\documentclass[12pt,preprint]{aastex}
\shorttitle{Seismic Analysis of 51 Peg}
\shortauthors{E. J. Murphy, P. Demarque, \& D. B. Guenther}

\begin{document}

\title{A Preliminary Seismic Analysis of 51 Peg: Large and 
Small Spacings from Standard Models}

\author{Eric J. Murphy , Pierre Demarque}
\affil{Department of Astronomy, Yale University, 
P.O. Box 208101, New Haven, CT 06520-8101}
\email{murphy@astro.yale.edu, demarque@astro.yale.edu}

\and

\author{D.B. Guenther}
\affil{Department of Astronomy and Physics, Saint Mary's University,
Halifax, N.~S., Canada, B3A~4R2}
\email{guenther@ap.stmarys.ca}

\begin{abstract}
  We present a preliminary theoretical seismic study of
  the astronomically famous star 51 Peg.  This is done by first
  performing 
  a detailed analysis within the Hertzsprung-Russell diagram (HRD).
  Using the Yale stellar evolution code (YREC), a grid of stellar
  evolutionary tracks has been constructed for the masses $1.00
  M_\odot, 1.05 M_\odot$ and $1.10 M_\odot$, in the metallicity range
  $Z=0.024-0.044$, and for values of the Galactic helium enrichment
  ratio (\(\Delta Y/ \Delta Z\)) in the range 0-2.5.  
  Along these evolutionary tracks, we select 75 stellar 
  model candidates that fall within the 51 Peg observational error box 
  in the HRD 
  (all turn out to have masses of $1.05 M_\odot$ and $1.10 M_\odot$).
  The corresponding allowable age range for these models, which 
  depends sensitively on the parameters of the model, is relatively 
  large and is \(\sim\)2.5 - 5.5 Gyr.
  For each
  of the 75 models, a non-radial pulsation analysis is carried out, and 
  the large and small frequency spacings are calculated. 
  The results show that just measuring the large and small 
  frequency spacings 
  will greatly reduce the 
  present uncertainties in the derived physical parameters and in 
  the age of 51 Peg.
  Finally we discuss briefly refinements in the physics of the models 
  and in the method of analysis which 
  will have to be included in future models to make the best of 
  the precise frequency determinations expected from space observations.

\end{abstract}

\keywords{stars: evolution - stars: fundamental parameters - stars:
  individual: 51 Peg - stars: oscillations}

\section{Introduction}
With current stellar structure and evolution theories, the placement
of a star on the Hertzsprung-Russell diagram (HRD) does not allow a
complete stellar parameterization.  This stems from the
fact that, except for the Sun, the number of free parameters is higher
than the number of observables.
We therefore look at seismology for additional observables to remove 
this degeneracy and to perform  
a more thorough analysis.   

Traditional observations of nearby stars lead to a range in
uncertainties for stellar radii and luminosity.  In the best cases an
accuracy of \(1\% - 10\%\) is attainable (Demarque et al. 1986).  
And with current modeling of stellar evolution, the stellar ages are 
not known better than \(\pm 10\%-25\%\).  
However, with the help of seismology, it is expected that the
uncertainty levels will be significantly reduced (Guenther 1998).

At the simplest level of detection, even 
when it is not feasible to determine individual frequencies, 
it is often possible to 
identify two quantities 
in the stellar {\it{p}}-mode oscillation
spectrum, the large and small frequency spacings (Tassoul 1980). 
These quantities are defined in {\S}3.1 below. 
As shown by Christensen-Dalsgaard (1984), 
these two quantities can be used to constrain the 
radius and age of the star, provided its chemical composition 
is known (Ulrich 1986).
Thus the large and small frequency spacings can provide 
important tests of the theory of stellar 
structure and evolution
for stars whose chemical composition and distance are known.    

We now take the case of the relatively well-studied 
nearby star 51 Peg and attempt
to constrain its basic parameters by combining stellar evolution 
modeling and stellar pulsation calculations.
51 Peg (HD 217014, HR 8729, HIP 113357) is classified as a G2.5IVa
(Hoffleit 1982).  It is best known for the discovery that 
it harbors an extra-solar, Jupiter-like planet 
(Mayor \& Queloz 1995; Marcy \& Butler 1998). 
  The excitement associated with the planetary system
of 51 Peg has placed it high on the target list for upcoming
space missions.  One mission of special interest is the 
recently launched Canadian MOST
(Microvariability and Oscillation of STars) microsatellite
which has begun observing a number of 
stars (Matthews
 et al. 2000), including 51 Peg.  We present this analysis in 
anticipation of the MOST observations.

This paper begins with the construction of stellar evolutionary tracks  
constrained
by the conventional astronomical parameters for 51 Peg.
Stellar models are then selected from these evolutionary tracks on 
the basis of whether or not they
fall within the observational error box mapped for 51 Peg onto the HRD.  
These
candidates are then pulsed and the calculated non-radial 
oscillation frequencies are used to derive both large and
small {\it{p}}-mode frequency spacings.  We show how, when combined with 
upcoming seismic data, the calculated large and small spacings will 
reduce the uncertainties in 51 Peg's physical parameters and age
estimate.  Thus, the spacings will help us refine future stellar
interior models for this star. 

The organization of the paper goes as follows:  In {\S}2.1 
we discuss the specifics
of our models and the physics implemented for their calculations.
Then in {\S}2.2 we outline how we go on to use observational data to
help constrain our models through our HRD analysis.  {\S}3 describes
the calculation of both large and small {\it{p}}-mode spacings and
how they can assist in constraining the parameters of 51 Peg.  
Lastly, in {\S}4, we 
summarize our analysis, which is necessarily 
exploratory in the absence of seismic observations. Some 
shortcomings of our models which will have to be addressed when
precise frequency measurements become available are discussed.  
Finally, we point out the need to devise more sophisticated 
approaches to analyze the high quality data expected from the space 
missions.

\section{Stellar Models}
\subsection{Model Physics}
All evolutionary tracks were computed using the Yale stellar
evolution code (YREC) (Guenther et al. 1992).  The
initial zero age main sequence (ZAMS) model used for 51 Peg was created
from pre-main sequence evolution calculations.  Post-main sequence
models of various compositions were then constructed by first rescaling
the composition of the ZAMS models.

In constructing the models we made use of the current OPAL equation
of state tables (Rogers et al.1996).  OPAL tables (Iglesias \& Rogers
1996) were also used to calculate interior opacities, while the
surface and atmosphere opacities were taken from the work of Alexander
and Ferguson (1994).  Due to the lack of better information, we assume
a solar mixture of heavy elements.  We accounted for the diffusion of
both the helium and heavy element abundances by weight (Y and Z,
respectively), using the diffusion  coefficients of Thoul et
al. (1994), in the way described by Guenther \& Demarque (1997) in
their models of the Sun.  

As expected, all the stellar models in this study exhibit 
a convective envelope.  In such stellar models, the 
calculated pulsational frequencies are known to be sensitive to the 
treatment of the atmosphere and outer convective layers 
(Guenther et al. 1992; Robinson et al. 2003). In the atmosphere, we
used the  Eddington T-$\tau$ relation for a grey atmosphere, and the  
mixing length theory to describe convectively unstable layers 
(B\"{o}hm-Vitense 1958). 
The calculated radius also depends on the choice of the 
mixing length parameter $\alpha$ (mixing length to scale-height 
ratio).  We set $\alpha$ = 1.7 for all models, close to the value 
required to reproduce the solar radius under the same physical 
assumptions and stellar evolution code. 
A fuller description of the treatment of the outer layers and of 
the constitutive physics incorporated into YREC can be found in
Guenther et al. (1992). 

Inspection of the evolutionary tracks in Figure~1 shows evidence in some 
of the tracks for a hydrogen exhaustion phase indicative of the 
presence of a convective core near the main sequence.
For simplicity, and because of the exploratory nature of this study, we 
have assumed in this study that no overshoot takes place at the edge
of convective cores, i.e. we have set \(\alpha_{OV}\) = 0.0, where
\(\alpha_{OV}\) is the core overshoot  
parameter.  The main consequence of core overshoot is to 
shorten the evolutionary lifetime.  Thus the 
presence of core overshoot 
would extend the core burning phase of evolution and 
increase the estimated age of 51 Peg (see e.g. 
Audard et al. 1995; Yi et al. 2000; Woo \& Demarque 2001). 
This important topic will be 
the subject of a separate study.

\subsection{Constraining Models with Observations}
Table~1 lists the principal constraints from observation.  They are
the position of 51 Peg in the HRD, with  the associated error box, and
two quantities derived from spectroscopy, i.e. the   
metallicity [Fe/H] and surface gravity $g$ (Santos et al. 2003).
The spectroscopic estimate of log~$g$ is a relatively weak constraint
which is easily satisfied since log~$g$ changes little all along the 
main sequence.

In mapping the observables onto the HRD, we have calculated the luminosity
of 51 Peg using the published visual magnitude in the Hipparcos
catalog (Perryman et al. 1997).  
Error estimates were derived based on the associated
Hipparcos parallax error.  Assuming a solar bolometric magnitude of
4.75 mag, we then calculated the bolometric correction by linear
interpolation in the color-correction tables of Green et
al. (1987) in \(T_{\rm eff}\), [Fe/H], and log {\it{g}}, to obtain our final
estimate of \(L_{\rm bol}\).  No transformation of \(T_{\rm eff}\) was
necessary since it maps directly from observation onto the theoretical 
HRD.  Thus
we constructed an error box based on 
 \begin{equation}
  L_{\rm bol} = 1.343_{-0.037}^{+0.025} L_{\sun}
 \end{equation}
and,
 \begin{equation}
  T_{\rm eff} = 5805 \pm 50{\rm K}
 \end{equation}
which is shown in Figure~1.
Where a star lies in the theoretical HRD (\(L_{\rm bol}\)
vs. \(T_{\rm eff}\)) 
is a complicated function of Y and Z, stellar mass, and age. 
In addition to the observational constraints from Table~1,
the mass and helium content of 51 Peg are only 
approximately known.  We used theoretical evolutionary tracks, 
as outlined in the next section, to 
better constrain these two parameters and to derive an age 
for 51 Peg.
Only those models which fell inside the error box in the HRD  
were considered as possible candidates for 51 Peg.

\subsection{Evolutionary Tracks and Candidates for Pulsation}
Table 2 lists the parameters adopted for the grid of evolutionary 
tracks.  The heavy element content by mass, Z, was derived from
[Fe/H].  The helium abundance Y, which cannot be observed directly,
was taken to be solar with several plausible  
values of the nucleosynthetic helium enrichment 
parameter \(\Delta Y/\Delta Z\) at higher Z.
The position in the HRD and the measured log$~g$ 
suggested a mass in the vicinity of one solar mass or slightly higher.  
Thus our grid of stellar evolutionary tracks was
constructed in the three parameter space of mass, Z, and \(\Delta
Y/\Delta Z\) given in Table 2.  The complete calculated grid
contained 45 evolutionary tracks of models for analysis.  Having 
constructed the grid we further selected those models which we considered
as candidates for pulsation on the basis of whether or not they landed
within the observational error box in the theoretical HRD.  
This error box is shown in Figure~1. A total of 75 models fall within
this region and are listed in Table 3.  

We see in Table 3 that none of the 1.0 $M_{\sun}$ evolutionary tracks 
pass through the error box in the HRD.
For the 1.05 $M_{\sun}$ tracks, only models for 
Z = 0.024 satisfy the HRD position constraint. In this case, 
the acceptable range in 
\(\Delta Y/\Delta Z\) is 0.5-2.5.  On the other hand, for the 
1.10 $M_{\sun}$ models, the range of possible metallicities includes 
Z = 0.034-0.044, with correspondingly higher helium abundances. 

We also note that on the basis of this analysis alone, the  
age of 51 Peg is still quite uncertain, 
and in the range \(\sim\)2.5 - 5.5 Gyr.  
For comparison, we recall   
previous age estimates of 8.5 Gyr, Noble et al. (2002), and 7.0 Gyr,
quoted by Edvardsson et al. (1993) and Ng \& Bertelli (1998).  All
these ages are based on sets of isochrones by VandenBerg (1985)
and Bertelli et al. (1994), and for various reasons, appear to be 
overestimates. More recently, Gonzalez (1998) estimates $6\pm 2$ Gyr 
for the age 
and $1.05 \pm 0.03 M_{\odot}$ for the mass of 51 Peg on the basis  
of the Schaller et al. (1992) and Schaerer et al. (1992) isochrones.  

In summary, for the purpose of this paper, and 
in view of the large uncertainties discussed above, 
we shall treat the 75 models listed in Table 3 as equally
likely model candidates for 51 Peg.
We shall see below that seismic 
observations of 51 Peg would greatly constrain parameters such as
mass,    age and 
metallicity. 

\section{Pulsation Analysis}
We now go
through the exercise of evaluating the {\it {p}}-mode frequencies 
of the 75 selected models using Guenther's
stellar pulsation code 
(Guenther 1994).
It is important to remember that the theoretical frequencies 
calculated here should 
not be expected to match the observed frequencies of 51 Peg  
very closely.  To start with, our theoretical models do not match 
either the radius or mass of 51 Peg precisely.  
But even in the hypothetical case in which the model radius 
and mass reproduce the stellar radius and mass with high 
precision (as in a calibrated 
standard solar models), the calculated frequencies could still differ 
appreciably from the observed frequencies due to the uncertainty in 
calculating the sound speed in the outer layers of the models where 
non-adiabatic effects become important.  
For instance, in 
the case of standard solar models constructed under the same physical 
assumptions as the 51 Peg models we present, this discrepancy can 
reach 10-20 $\mu Hz$ at the higher frequencies (see Guenther \& Demarque 1997; 
Winnick et al. 2001).  This discrepancy is reduced by a factor of four in 
standard solar models that include a more realistic description of 
convection in the atmosphere (Li et al. 2002).   
  Because the 
discrepancies are themselves a function of frequency, these 
could also result in a smaller but significant 
effect on the calculated  
large spacings discussed in \S 3.2 and \S 4. 

On the other hand, the small spacings, being primarily sensitive 
to the central 
concentration of the model, are less affected by the details of the 
surface boundary conditions. The small spacings are 
sensitive primarily to age and to the size of the convective 
core when one is present.

\subsection{Generating $p$-Mode Frequencies and Spacings}

Guenther's (1994) pulsation code allows for the non-adiabatic 
corrections in the pulsation needed
to account for radiation in the Eddington approximation.  The accuracy 
of these corrections is a function of frequency such that at high frequencies
it is harder to account for non-adiabatic effects due to
convection-oscillation interactions and thus errors are larger.  
We define the large 
spacings \(\Delta \nu\) and small
spacings \(\delta \nu\) in the usual way, such that,
\begin{equation}
  \Delta \nu_{\it{n,l}} \equiv \nu(n,l) - \nu(n-1,l)
\end{equation}
and,
\begin{equation}
  \delta \nu_{\it{n,l}} \equiv \nu(n,l) - \nu(n-1,l+2),
\end{equation}
where {\it{n}} is the radial order, {\it{l}} is the azimuthal order,
and \(\nu\) is the frequency.  The importance of these spacings and
association with upcoming observations is discussed in \S 3.2 and \S
3.3. 

Large and small spacings were obtained
over the ranges of \(n = 1-30\) and \(l = 0-3\).  These values of
{\it{n}} and {\it{l}} correspond to a frequency range of \(\sim 1000
\mu\)Hz to \(\sim 4000\mu\)Hz.  We, however, only plot the
frequency range out to \(\sim 3500\mu\)Hz due to the expected 
acoustic cutoff just beyond this limit.  

In Table 3 we give the parameters
of each model 
along with characteristic large and small spacings. The mean large 
spacings are averages over $\ell$ = 0,1,2,3 and $n$ = 10,11,12,...,30.
The mean small spacings are averages over $n$ = 10,11,12,...,30 at a 
fixed $\ell$.  For illustration, we have plotted in Figures 2 and 3, 
average large and
small spacings respectively vs. frequency for a given subset of
nine individual stellar models along the evolutionary track  
having M \(= 1.05M_{\sun}\), X \(=
0.686\), and Z \(= 0.024\), which fell within our error box.

\subsection{Large Spacings}
It is well known from
asymptotic theory that the large spacings are mainly sensitive to the
stellar radius (Tassoul 1980; Christensen-Dalsgaard 1984).  
More precisely, the asymptotic behavior of \(\Delta
\nu\) is expected to scale with \((M/R^{3})^{1/2}\) where {\it{M}} is
the mass of the star and {\it{R}} is its radius.  It is therefore not
surprising that perturbations in Y and Z will have less effect
on the large spacings than uncertainties in the luminosity and \(T_{\rm
  eff}\) \((L \propto R^{2}T_{\rm eff}^{4}\)).   

As pointed out by Guenther (1998), one of the most easily identifiable
characteristics in the {\it{p}}-mode spectrum are the large spacings.
They are seen as a peak in the Fourier transform of the power
spectrum and because they are mostly uncontaminated by composition 
effects, these large separations provide a much
more precise way to measure stellar radii compared with conventional
techniques.  In Figure 4 we plot the 
average large spacings vs. model radius.

It is clear from the plot that radius increases as \(\Delta \nu\)
decreases with very little scatter for the 1.05 \(M_{\sun}\) models.  We
find that a degeneracy in predicted radius occurs for models of
different mass.  Specifically, in our analysis, we see that \(\Delta \nu\)
calculated from 1.10\(M_{\sun}\) models lie on a nearly parallel line
to that generated by the 1.05\(M_{\sun}\) models
having a vertical shift of around 2.7\(\mu\)Hz in \(\Delta \nu\) and a
much larger scatter.  Such a shift implies that an observed \(\Delta
\nu\) can lead to the calculation of a radius differing by nearly
\(\sim .007R_{\sun}\) depending on whether the stellar mass is 1.05
\(M_{\sun}\) or 1.10 \(M_{\sun}\).  This degeneracy may be lifted by
using solar models with the assumption of a homologous scaling.

It is sometimes convenient to assume homology to 
compare theoretical models  
by introducing a ``reduced'' radius (see e.g. Fernandes and Monteiro
2003) such that,
\begin{equation}
\Delta \nu_{r} = \Delta \nu_{n,l} \left(\frac{R}{R_{\sun}}\right)^{3/2},
\end{equation}
  We have listed the values of \(\Delta \nu_{r}\) for each
model in Table 3.

It is easily seen that the values of the ``reduced'' spacings are
relatively consistent for each mass such that \(\Delta \nu_{r} \sim
141\mu\)Hz for 1.05 \(M_{\sun}\) models and \(\Delta \nu_{r} \sim
144\mu\)Hz for 1.10 \(M_{\sun}\).  Were our stellar models purely
homologous, then the reduced spacing would only be a function of
mass.  This would then lead us to calculate the correct mass when
\(\Delta \nu_{r} = \Delta \nu_{\sun}\).  We
could therefore obtain a correct radius for 51 Peg from the \(\Delta
\nu\) versus radius plot.  Note however that this is only approximately
true.  Scatter exists within the calculated \(\Delta \nu_{r}\) due to
significant departure from homology between the solar interior
structure and the models for 51 Peg.  These fluctuations in \(\Delta
\nu_{r}\) are due to variations in chemical composition and age, as
well as other input physics of each model such as the depth of the
convection zone.  Direct measurements of stellar diameters from
interferometric observations may be able to help provide an
independent check with these asteroseismic predictions (Boden et al.
1998; Kervella et al.  2003a,b). 

\subsection{Small Spacings}
As mentioned earlier, it was initially believed 
that the calculation of small spacings could
put a constraint on the age of the star (Christensen-Dalsgaard 1984).  
However, it was subsequently 
realized that only if the
composition of the star is known completely can one use the small
spacings to correctly identify a stellar age unequivocally (Ulrich 1986).  
This point is illustrated in Figure 5, where we plot the 
average small spacings  
vs. the calculated age for all potential 51 Peg model candidates. 
  It is obvious that the models of different
chemical compositions each define a distinct locus of points which
seems to follow offset linear relationships such that age increases
with decreasing \(\delta \nu\). Each locus of model points at each
composition seem to have very nearly the same slope, but shifted both
vertically and horizontally. Thus, the various chemical compositions
create a degeneracy in age determination.  

The small spacings, like the large spacings, will be visible as peaks
in the Fourier transform of the power spectrum.  However, as Figure 5
clearly shows, the small spacings seem rather sensitive to
composition and therefore to the structure of the core.  The extreme
sensitivity of the stellar core density stratification to several  
parameters has been discussed in the case of 
$\alpha$ Cen~A (Kim 1999; Guenther \& Demarque 2000; Morel et al. 2000).
Though sensitive to composition, the small spacings can still be used to 
further constrain the models and obtain a radius once precise observations
are available.  The measurement of the small spacing will enable us to
constrain the mass in Figure 5 to a much narrower range than is 
presently possible on the basis of HRD position alone.  Note that  
there is an extremely small overlap in
\(\delta \nu\) for stars with 1.05 \(M_{\sun}\) and 1.10 \(M_{\sun}\).
Given an observed small spacing, one could determine 
more precisely which range of models to use to calculate the
radius in Figure 4.  Thus, by obtaining both small and large spacings
it is possible in principle to reduce considerably  the \(\Delta \nu\)-radius 
degeneracy with mass, and to reduce the age uncertainty to less than 1.0 Gyr.

\section{Summary and conclusions}
Using the constraints from the best available observations of the 
star's position in the 
HRD, as well as the available spectroscopic constraints 
on its metallicity and surface gravity, 
we have made an analysis of the pulsation properties of a set of  
evolutionary models for the star 51 Peg.  All of these models were 
selected from the evolutionary tracks 
for their location within the HRD error box of 51 Peg. 
These evolutionary tracks covered a range of metallicities. 
(Note that spectroscopy also 
provided an estimate of the surface gravity, a weaker constraint, 
but a useful consistency check in this case). 
In order to 
construct the tracks, two additional parameters had 
to be chosen , the helium abundance, and the mass.
The adopted composition and mass parameters for the 45 
evolutionary tracks are given in Table 2.

Table 3 lists the 
acceptable models for 51 Peg. The
range in stellar parameter space through the initial survey 
within the HRD is relatively large.  
In this case, the conventional astronomical tests 
 leave a mass uncertainty of about \(5\%\) and a large range in age 
\((2.5 - 5.5 Gyr\)).

Thus it is clear that {\it{p}}-mode observations, 
even at the most basic level, will aid in the
analysis of stellar evolution and structure.  We have outlined the
usefulness of both the calculation of the large and small spacings as
well as discuss how both can be used in conjunction with one another to
lift existing degeneracies in detailed stellar parameterization.
In particular,
since the large {\it{p}}-mode spacings are primarily sensitive to radius while
insensitive to composition, they will, along with precise parallax
measurements, provide improvements in the calculation of stellar 
radii and \(T_{\rm eff}\). And, as a result, it should be 
possible to reduce the uncertainty in 
age to less than 1 Gyr.

We emphasize that the expected high quality 
and quantity of the space observations, including the precise 
measurements of individual oscillation mode frequencies, 
will necessitate more sophisticated stellar models than presented in this paper.
As pointed out by Guenther \& Demarque (2000) in the case of 
$\alpha$ Cen~A, 
quality seismic data can reveal the presence of a convective core, either 
at this time or during the 
past evolution of the star, and the size of the mixed region. 
This important point has recently been made more graphically by 
Guenther \& Brown (2004).
Observations 
from space will be needed to settle this issue. 
The recent ground-based observations (Bouchy \& Carrier 2002; 
Carrier \& Bourban 2003) of individual frequencies 
in $\alpha$ Cen~A and B have 
improved our knowledge of the system, but the uncertainties are 
still such that interior models 
constructed with those data have yet to reach a consensus (Th\'{e}venin 
et al. 2002; Thoul et al. 2003).   

Two significant effects with seismic signatures 
should be included in future models 
of 51 Peg, i.e. convective core overshoot, and turbulence in the outer layers 
of the convection zones.  The extent of core overshoot may be 
detectable from the 
small frequency spacings and from the signature of mode bumping.
The detailed comparison of individual mode frequencies will also require 
taking into account the effects of turbulence in the outer convectively 
unstable layers in the stellar models which shift the observed frequencies.  
The parameterization 
of turbulence tested on the 
Sun by Li et al. (2002) can be applied to models for Sun-like stars 
as well.  This parameterization can be extended to other stars by 
using the three-dimensional 
radiative hydrodynamical simulations of Robinson et al. (2003), which are 
based on the same microscopic physics, and can readily be parameterized  
 in the YREC stellar evolution code. 
We intend to explore both the effects of convective core overshoot and of
including turbulence in modeling the outer layers in subsequent papers.
 We now
await the first results from MOST on 51 Peg to 
compare our theory to observations.

At the same time, high quality observations from space will also 
require more powerful techniques of analysis than we have used  
to interpret the observations and take advantage of the 
wealth of information contained in a spectrum of well
determined frequencies (Brown et al. 1984).  Powerful new 
approaches are being developed to extract precise masses and
metallicities (Guenther \& Brown 2004) and 
envelope helium abundances  (Basu et al. 2003) for individual stars from 
precisely measured pulsation frequencies. 

\acknowledgements

We would like to thank Frederic Th\'{e}venin for his careful
refereeing of this paper.  This research has been supported in part by
NASA grants NAG5-8406 and NAG5-13299 (PD).  DBG acknowledges support
from an operating research grant from NSERC.

\clearpage
\begin{figure}
\plottwo{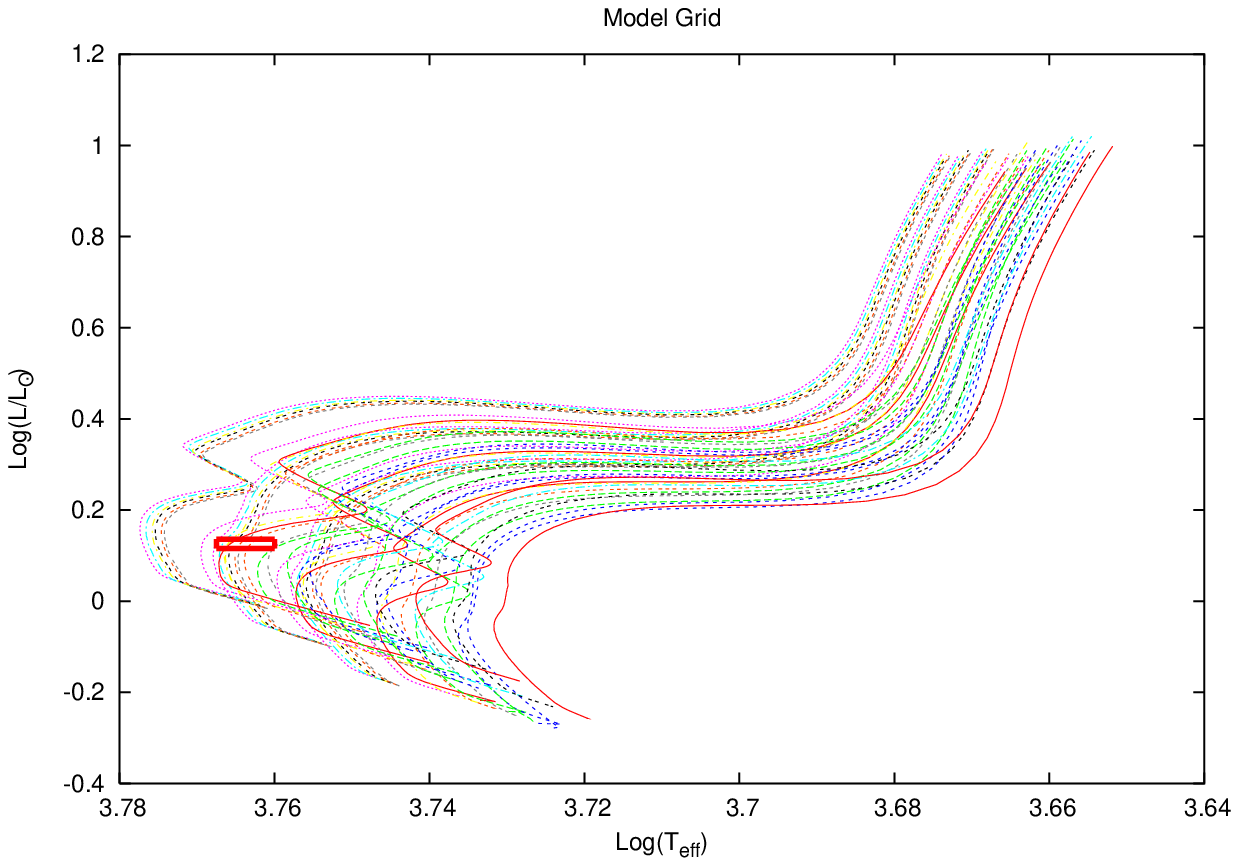}{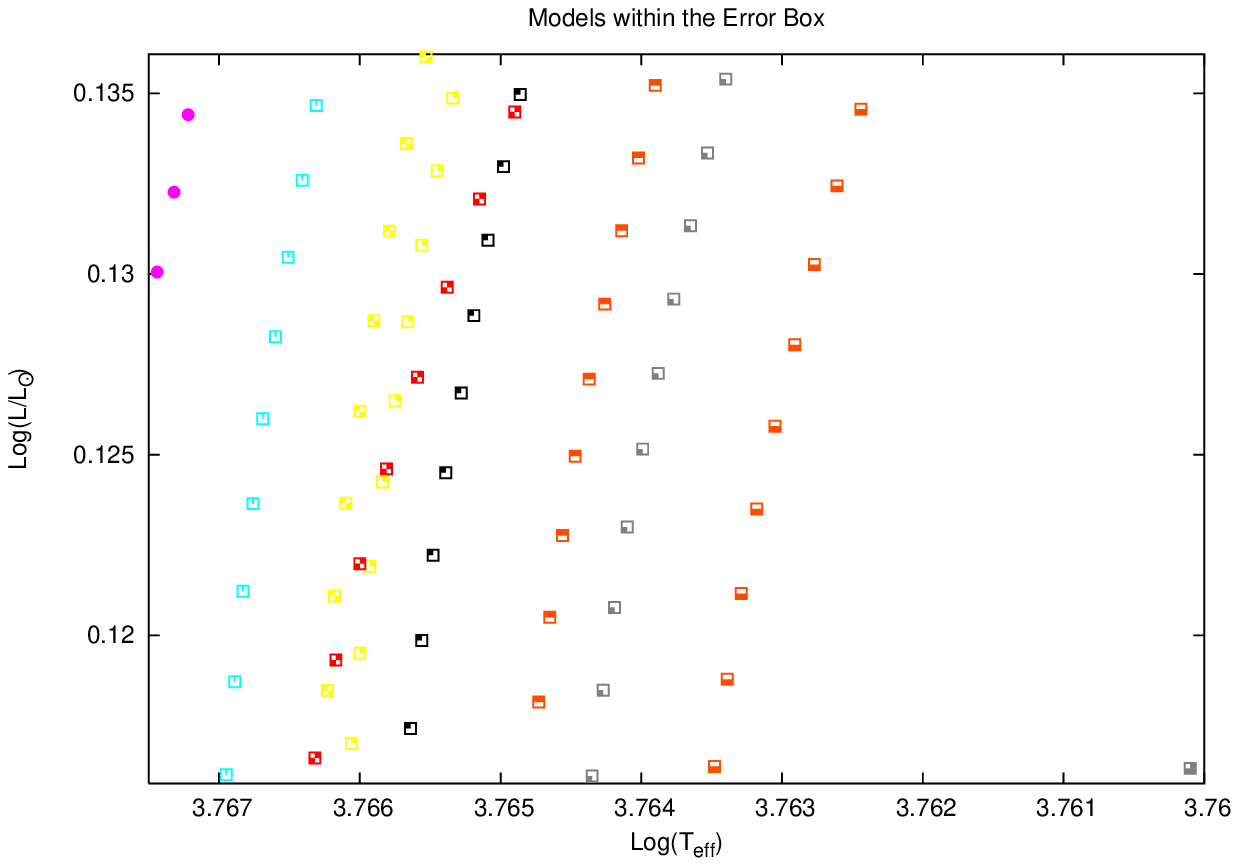}
\caption{To the left, we plot our initial grid of evolutionary 
  tracks (45 in total) in the theoretical HRD, 
  spanning the parameter values listed in Table 2.  The observational
  error box is drawn.  To the right, 
  we plot a blow-up of the error box showing the 75 acceptable 
  individual stellar models along each track. Parameters of 
  the models are listed in Table 3.
  \label{fig1}} 
\end{figure}

\clearpage
\begin{figure}
\plotone{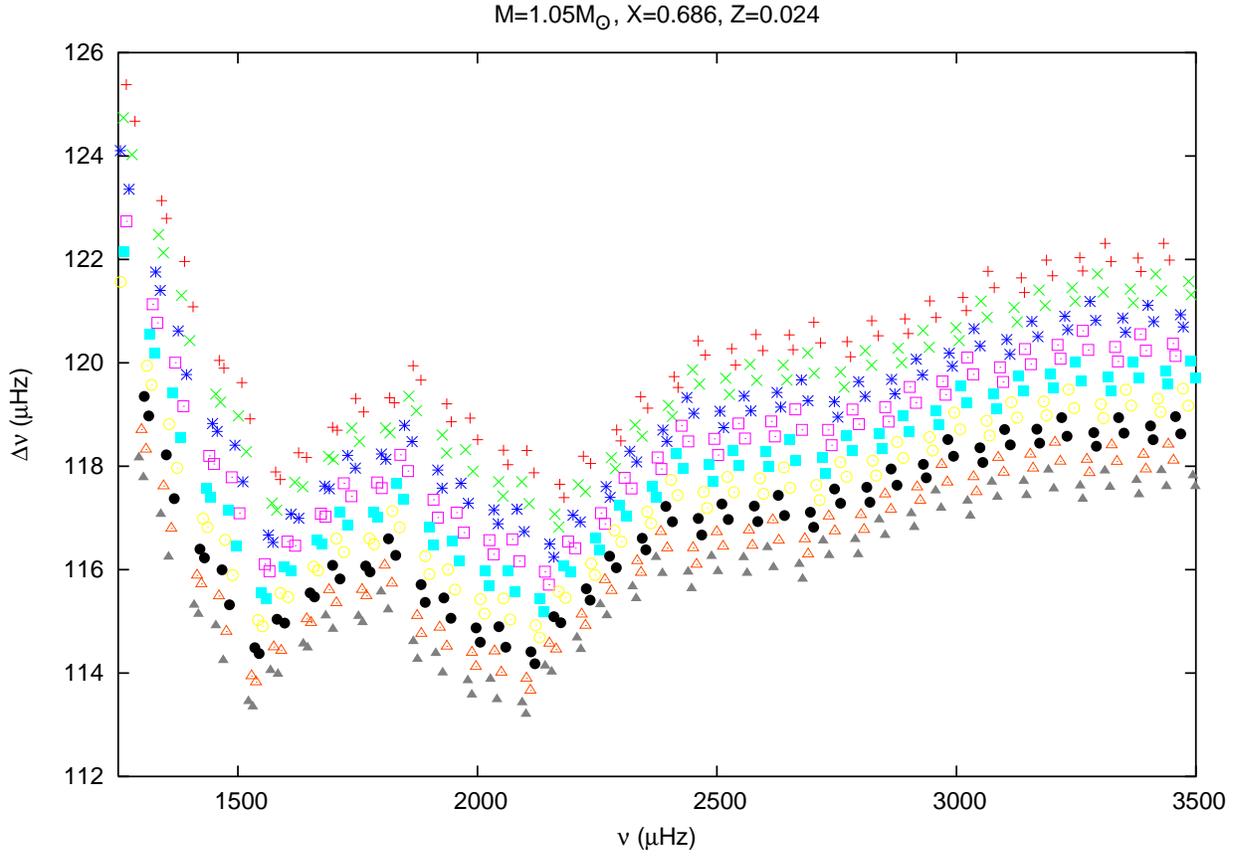}
\caption{Large frequency spacings vs. frequency for 9 individual stellar 
  models (models 4-12 in Table 3) along the evolutionary track 
  with parameters: $M = 1.05M_{\sun}$, $X =0.686$, and $Z =
  0.024$. \label{fig2}} 
\end{figure}

\clearpage
\begin{figure}
\plotone{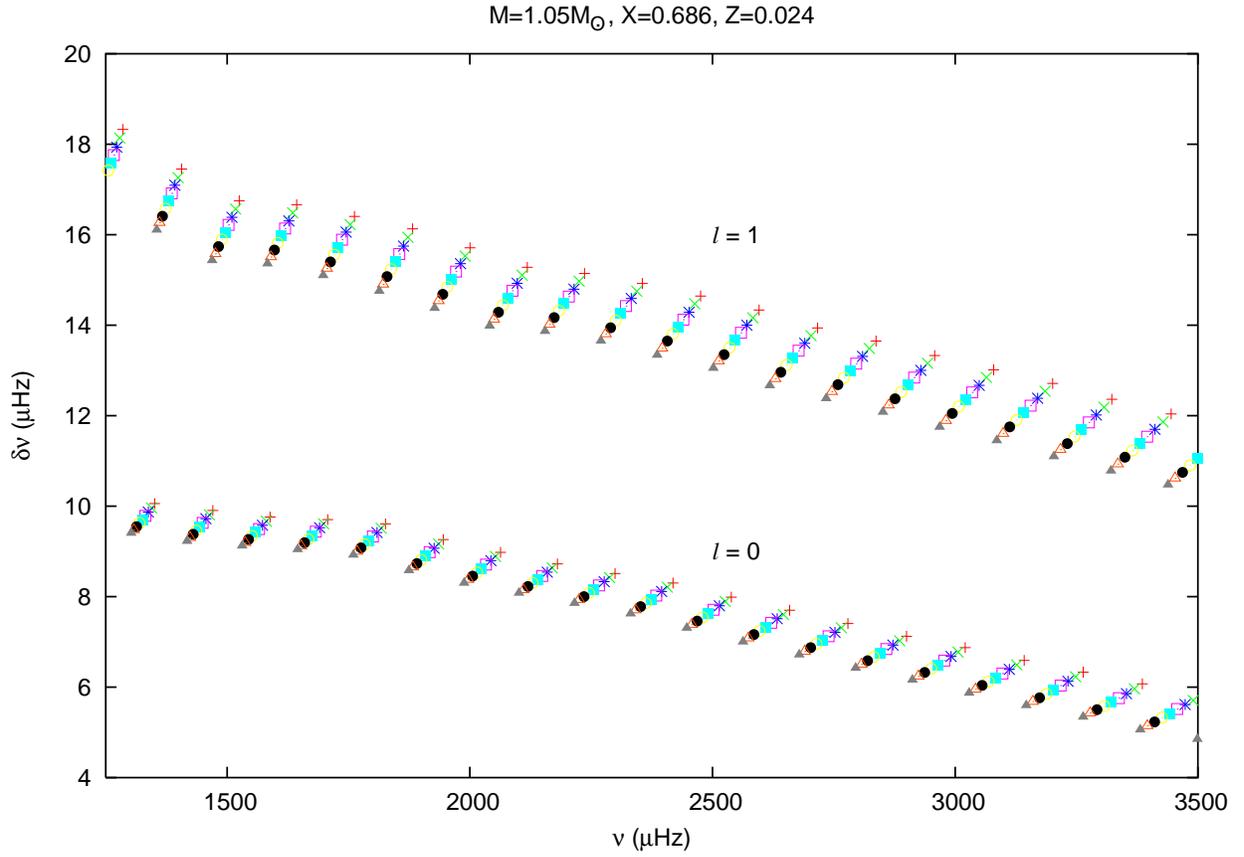}
\caption{Small frequency spacings vs. frequency for the same 9 stellar 
  models as in Fig.2. Spacings for $\ell = 0$ and $\ell = 1$ are marked.
  \label{fig3}} 
\end{figure}

\clearpage
\begin{figure}
\plotone{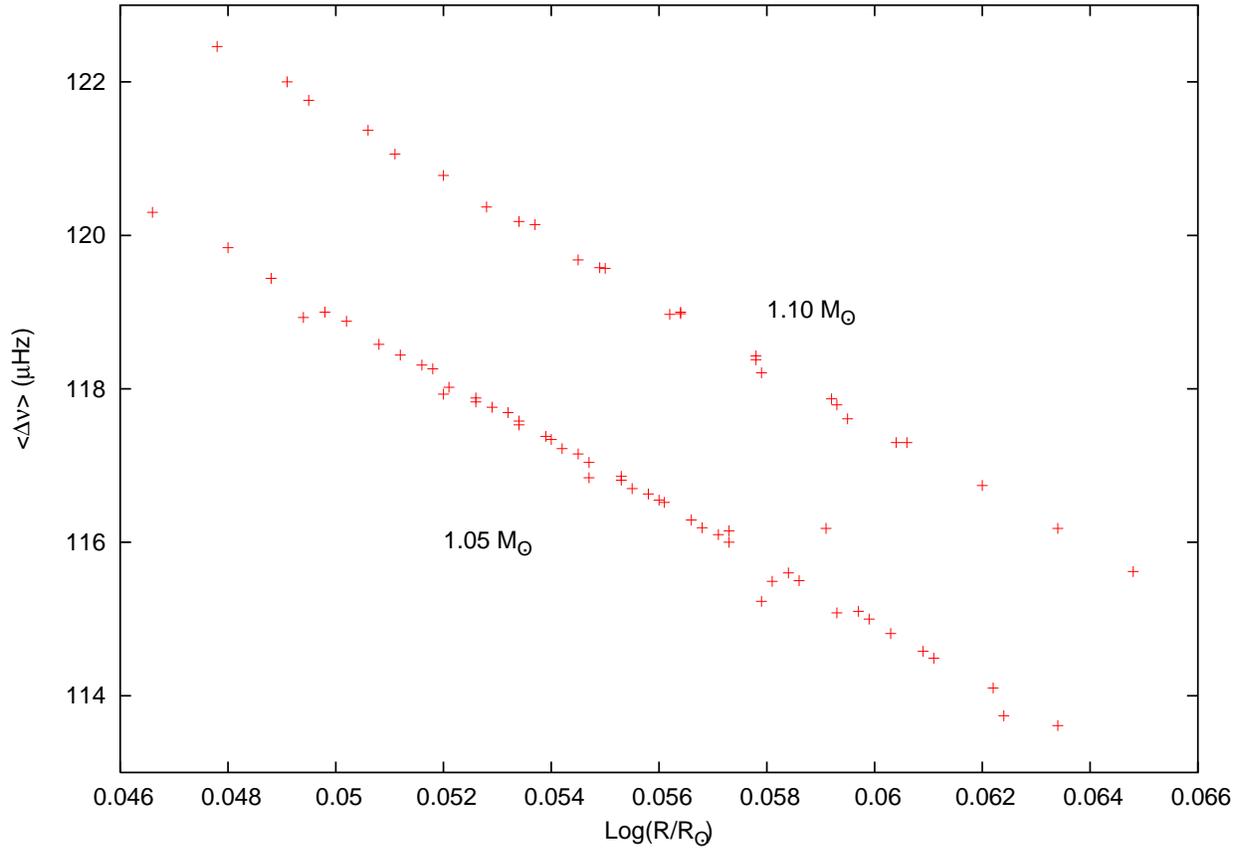}
\caption{Average calculated large frequency spacings vs. stellar radius 
  (in solar units) for
  each of the 75 stellar models.  \label{fig4}}
\end{figure}

\clearpage
\begin{figure}
\plottwo{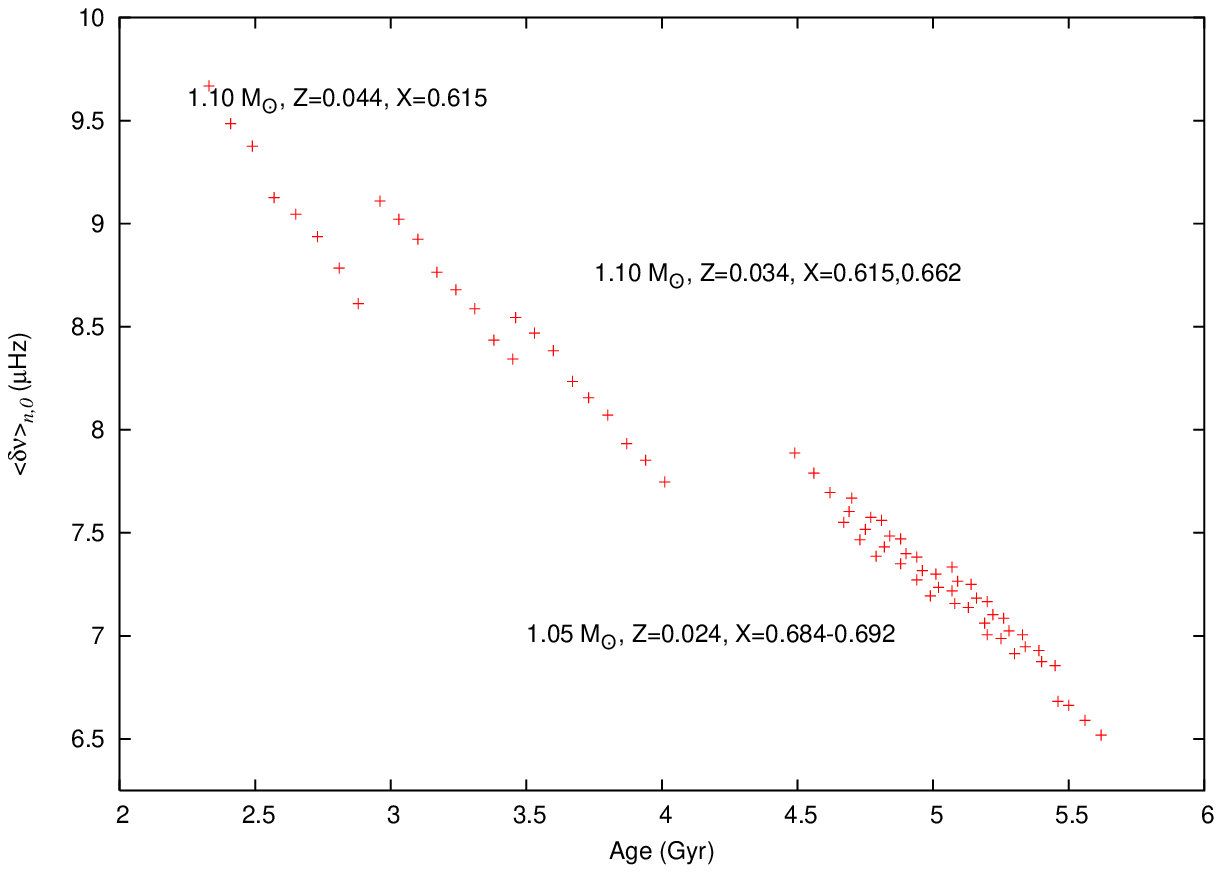}{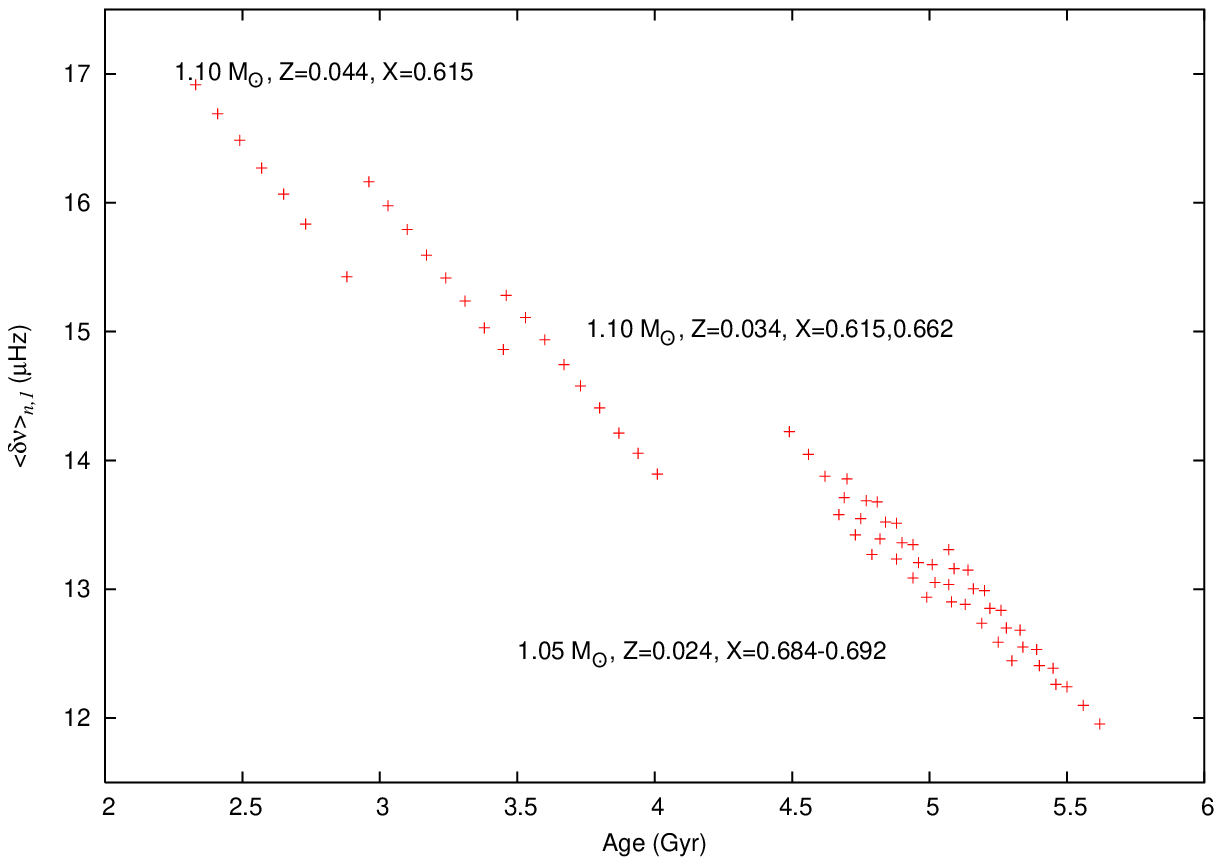}
\caption{Average calculated small frequency spacings vs. age for
  each of the 75 stellar models. The left-hand panel shows the spacings 
  for $\ell = 0$, 
  the right-hand panel for $\ell = 1$. \label{fig5}}
\end{figure}

\clearpage
\begin{deluxetable}{crr}
\tablecaption{Observational Data of 51 Peg\label{tbl-1}}
\tablewidth{0pt}
\tablehead{
\colhead{Observable} & \colhead{Value} & \colhead{Source}
}
\startdata
$T_{\rm eff}$  & 5805 $\pm$ 50K & Santos et al. 2003 \\
log {\it{g}} & 4.51 $\pm$ 0.15 dex & Santos et al. 2003 \\
$\pi$ (\arcsec) & 0.06510 $\pm$ 0.00076 & {\it{Hipparcos}} Catalog \\
$[$Fe/H$]$ & 0.21 $\pm$ 0.06 dex & Santos et al. 2003 \\
\enddata
\end{deluxetable}

\clearpage
\begin{deluxetable}{crrr}
\tablecaption{Input Parameters for Model Tracks \label{tbl-2}}
\tablewidth{0pt}
\tablehead{
\colhead{Variable} & \colhead{minimum value} & \colhead{maximum
  value} & \colhead{$\delta$}
}
\startdata
$M/M_{\sun}$ & 1.00 & 1.10 & 0.05 \\
Z & 0.024 & 0.044 & 0.01 \\ 
$\Delta$Y/$\Delta$Z & 0.0 & 2.5 & 0.5 \\
\enddata
\tablecomments{$\delta$ defines the increment between minimum and
  maximum parameter values used to create the model array.}
\end{deluxetable}

\clearpage
\begin{deluxetable}{ccccccccccccc}
\tabletypesize{\scriptsize}
\tablecaption{Model Parameters \label{tbl-4}}
\tablewidth{0pt}
\tablehead{
\colhead{Star} & \colhead{\it M}   & \colhead{\it X} 
&\colhead{$\Delta$Y/$\Delta$Z}  &
\colhead{\it Z} & \colhead{Log {\it T}$_{\rm eff}$}  & \colhead{Log {\it L}}&
\colhead{Log {\it R}}& 
\colhead{Age} & \colhead{$<$\(\Delta\nu\)$>$} & 
\colhead{$<$\(\delta\nu\)$>$} & \colhead{$<$\(\delta\nu\)$>$} & 
\colhead{$<$\(\Delta\nu_{r}\)$>$} \\
\colhead{}&\colhead{($M_{\sun}$)}&\colhead{}&\colhead{}&\colhead{}&\colhead{}
&\colhead{($L_{\sun}$)}& \colhead{($R_{\sun}$)}&\colhead{(Gyr)}&
\colhead{}&\colhead{({\it n,0})}&\colhead{({\it n,1})} & 
\colhead{}
}
\startdata
1   &1.05   &0.684   &2.50   &0.0240     &3.7675     &0.128         &0.0526       &4.67      &117.83    &7.5502  &13.579  &141.33    \\
2   &1.05   &0.684   &2.50   &0.0240     &3.7674     &0.130         &0.0539       &4.73      &117.38    &7.4667  &13.422  &141.42    \\
3   &1.05   &0.684   &2.50   &0.0240     &3.7673     &0.132         &0.0553       &4.79      &116.86    &7.3860  &13.269  &141.44    \\
4   &1.05   &0.686   &2.00   &0.0240     &3.7670     &0.113         &0.0466       &4.49      &120.30    &7.8876  &14.224  &141.29    \\
5   &1.05   &0.686   &2.00   &0.0240     &3.7670     &0.116         &0.0480       &4.56      &119.84    &7.7898  &14.047  &141.43    \\
6   &1.05   &0.686   &2.00   &0.0240     &3.7669     &0.119         &0.0494       &4.62      &118.93    &7.6952  &13.878  &141.05    \\
7   &1.05   &0.686   &2.00   &0.0240     &3.7668     &0.121         &0.0508       &4.69      &118.58    &7.6040  &13.711  &141.30    \\
8   &1.05   &0.686   &2.00   &0.0240     &3.7668     &0.124         &0.0521       &4.75      &118.02    &7.5165  &13.549  &141.29    \\
9   &1.05   &0.686   &2.00   &0.0240     &3.7667     &0.126         &0.0534       &4.82      &117.53    &7.4320  &13.391  &141.34    \\
10  &1.05   &0.686   &2.00   &0.0240     &3.7666     &0.128         &0.0547       &4.88      &116.84    &7.3498  &13.235  &141.16    \\
11  &1.05   &0.686   &2.00   &0.0240     &3.7665     &0.130         &0.0560       &4.94      &116.55    &7.2716  &13.087  &141.43    \\
12  &1.05   &0.686   &2.00   &0.0240     &3.7664     &0.133         &0.0573       &4.99      &116.15    &7.1941  &12.938  &141.57    \\
13  &1.05   &0.688   &1.50   &0.0240     &3.7661     &0.114         &0.0488       &4.70      &119.44    &7.6681  &13.857  &141.36    \\
14  &1.05   &0.688   &1.50   &0.0240     &3.7661     &0.117         &0.0502       &4.77      &118.88    &7.5753  &13.688  &141.38    \\
15  &1.05   &0.688   &1.50   &0.0240     &3.7660     &0.120         &0.0516       &4.84      &118.31    &7.4847  &13.522  &141.37    \\
16  &1.05   &0.688   &1.50   &0.0240     &3.7659     &0.122         &0.0529       &4.90      &117.76    &7.3989  &13.361  &141.36    \\
17  &1.05   &0.688   &1.50   &0.0240     &3.7658     &0.124         &0.0542       &4.96      &117.22    &7.3162  &13.206  &141.38    \\
18  &1.05   &0.688   &1.50   &0.0240     &3.7658     &0.126         &0.0555       &5.02      &116.70    &7.2359  &13.053  &141.38    \\
19  &1.05   &0.688   &1.50   &0.0240     &3.7657     &0.129         &0.0568       &5.08      &116.19    &7.1570  &12.902  &141.38    \\
20  &1.05   &0.688   &1.50   &0.0240     &3.7656     &0.131         &0.0581       &5.14      &115.49    &\nodata &\nodata &141.14    \\
21  &1.05   &0.688   &1.50   &0.0240     &3.7654     &0.133         &0.0593       &5.20      &115.08    &7.0054  &\nodata &141.25    \\
22  &1.05   &0.689   &1.00   &0.0240     &3.7657     &0.115         &0.0498       &4.81      &119.00    &7.5607  &13.678  &141.36    \\
23  &1.05   &0.689   &1.00   &0.0240     &3.7656     &0.117         &0.0512       &4.88      &118.44    &7.4708  &13.512  &141.37    \\
24  &1.05   &0.689   &1.00   &0.0240     &3.7656     &0.120         &0.0526       &4.94      &117.88    &7.3822  &13.347  &141.37    \\
25  &1.05   &0.689   &1.00   &0.0240     &3.7655     &0.122         &0.0540       &5.01      &117.34    &7.2997  &13.192  &141.38    \\
26  &1.05   &0.689   &1.00   &0.0240     &3.7654     &0.124         &0.0553       &5.07      &116.81    &7.2187  &13.037  &141.39    \\
27  &1.05   &0.689   &1.00   &0.0240     &3.7653     &0.127         &0.0566       &5.13      &116.29    &7.1378  &12.883  &141.39    \\
28  &1.05   &0.689   &1.00   &0.0240     &3.7652     &0.129         &0.0579       &5.19      &115.23    &7.0621  &12.736  &140.72    \\
29  &1.05   &0.689   &1.00   &0.0240     &3.7651     &0.131         &0.0591       &5.25      &116.18    &6.9868  &12.589  &142.48    \\
30  &1.05   &0.689   &1.00   &0.0240     &3.7650     &0.133         &0.0603       &5.30      &114.81    &6.9136  &12.445  &141.41    \\
31  &1.05   &0.691   &0.50  &0.0240     &3.7648     &0.116         &0.0520       &5.03      &117.93    &\nodata &\nodata &141.16    \\
32  &1.05   &0.691   &0.50  &0.0240     &3.7647     &0.118         &0.0534       &5.09      &117.58    &7.2654  &13.160  &141.40    \\
33  &1.05   &0.691   &0.50  &0.0240     &3.7647     &0.120         &0.0547       &5.16      &117.04    &7.1834  &13.004  &141.40    \\
34  &1.05   &0.691   &0.50  &0.0240     &3.7646     &0.123         &0.0561       &5.22      &116.52    &7.1034  &12.852  &141.41    \\
35  &1.05   &0.691   &0.50  &0.0240     &3.7645     &0.125         &0.0573       &5.28      &116.00    &7.0238  &12.699  &141.41    \\
36  &1.05   &0.691   &0.50  &0.0240     &3.7644     &0.127         &0.0586       &5.34      &115.50    &6.9478  &12.551  &141.41    \\
37  &1.05   &0.691   &0.50  &0.0240     &3.7643     &0.129         &0.0599       &5.40      &115.00    &6.8747  &12.407  &141.42    \\
38  &1.05   &0.691   &0.50  &0.0240     &3.7641     &0.131         &0.0611       &5.46      &114.49    &6.6822  &12.261  &141.40    \\
39  &1.05   &0.691   &0.50  &0.0240     &3.7640     &0.133         &0.0624       &5.51      &113.74    &\nodata &\nodata &141.07    \\
40  &1.05   &0.692   &0.50  &0.0240     &3.7644     &0.114         &0.0518       &5.07      &118.26    &7.3337  &13.307  &141.40    \\
41  &1.05   &0.692   &0.50  &0.0240     &3.7643     &0.116         &0.0532       &5.14      &117.69    &7.2495  &13.148  &141.41    \\
42  &1.05   &0.692   &0.50  &0.0240     &3.7643     &0.118         &0.0545       &5.20      &117.15    &7.1661  &12.989  &141.41    \\
43  &1.05   &0.692   &0.50  &0.0240     &3.7642     &0.121         &0.0558       &5.26      &116.63    &7.0860  &12.835  &141.42    \\
44  &1.05   &0.692   &0.50  &0.0240     &3.7641     &0.123         &0.0571       &5.33      &116.10    &7.0058  &12.682  &141.42    \\
45  &1.05   &0.692   &0.50  &0.0240     &3.7640     &0.125         &0.0584       &5.39      &115.60    &6.9293  &12.533  &141.42    \\
46  &1.05   &0.692   &0.50  &0.0240     &3.7639     &0.127         &0.0597       &5.45      &115.10    &6.8553  &12.387  &141.44    \\
47  &1.05   &0.692   &0.50  &0.0240     &3.7638     &0.129         &0.0609       &5.50      &114.58    &6.6625  &12.242  &141.42    \\
48  &1.05   &0.692   &0.50  &0.0240     &3.7637     &0.131         &0.0622       &5.56      &114.10    &6.5905  &12.098  &141.42    \\
49  &1.05   &0.692   &0.50  &0.0240     &3.7635     &0.133         &0.0634       &5.62      &113.61    &6.5189  &11.954  &141.43    \\
50  &1.10   &0.615   &2.50   &0.0440     &3.7665     &0.114         &0.0478       &2.33      &122.46    &9.6681  &16.917  &144.45    \\
51  &1.10   &0.615   &2.50   &0.0440     &3.7663     &0.117         &0.0495       &2.41      &121.76    &9.4855  &16.692  &144.44    \\
52  &1.10   &0.615   &2.50   &0.0440     &3.7662     &0.119         &0.0511       &2.49      &121.06    &9.3758  &16.485  &144.44    \\
53  &1.10   &0.615   &2.50   &0.0440     &3.7660     &0.122         &0.0528       &2.57      &120.37    &9.1264  &16.270  &144.45    \\
54  &1.10   &0.615   &2.50   &0.0440     &3.7658     &0.125         &0.0545       &2.65      &119.68    &9.0453  &16.067  &144.45    \\
55  &1.10   &0.615   &2.50   &0.0440     &3.7656     &0.127         &0.0562       &2.73      &118.97    &8.9367  &15.834  &144.45    \\
56  &1.10   &0.615   &2.50   &0.0440     &3.7654     &0.130         &0.0579       &2.81      &118.21    &8.7847  &\nodata &144.36    \\
57  &1.10   &0.615   &2.50   &0.0440     &3.7652     &0.132         &0.0595       &2.88      &117.61    &8.6115  &15.426  &144.46    \\
58  &1.10   &0.656   &2.00   &0.0340     &3.7663     &0.116         &0.0491       &2.96      &122.00    &9.1095  &16.163  &144.53    \\
59  &1.10   &0.656   &2.00   &0.0340     &3.7662     &0.118         &0.0506       &3.03      &121.37    &9.0214  &15.977  &144.54    \\
60  &1.10   &0.656   &2.00   &0.0340     &3.7662     &0.121         &0.0520       &3.10      &120.78    &8.9242  &15.793  &144.54    \\
61  &1.10   &0.656   &2.00   &0.0340     &3.7661     &0.124         &0.0534       &3.17      &120.18    &8.7645  &15.593  &144.54    \\
62  &1.10   &0.656   &2.00   &0.0340     &3.7660     &0.126         &0.0549       &3.24      &119.58    &8.6796  &15.416  &144.54    \\
63  &1.10   &0.656   &2.00   &0.0340     &3.7659     &0.129         &0.0564       &3.31      &118.98    &8.5877  &15.237  &144.55    \\
64  &1.10   &0.656   &2.00   &0.0340     &3.7658     &0.131         &0.0578       &3.38      &118.38    &8.4349  &15.030  &144.54    \\
65  &1.10   &0.656   &2.00   &0.0340     &3.7657     &0.134         &0.0593       &3.45      &117.79    &8.3435  &14.861  &144.55    \\
66  &1.10   &0.662   &1.50   &0.0340     &3.7635     &0.114         &0.0537       &3.46      &120.14    &8.5445  &15.281  &144.60    \\
67  &1.10   &0.662   &1.50   &0.0340     &3.7635     &0.116         &0.0550       &3.53      &119.57    &8.4686  &15.109  &144.60    \\
68  &1.10   &0.662   &1.50   &0.0340     &3.7634     &0.119         &0.0564       &3.60      &119.00    &8.3837  &14.936  &144.61    \\
69  &1.10   &0.662   &1.50   &0.0340     &3.7633     &0.121         &0.0578       &3.67      &118.43    &8.2343  &14.744  &144.60    \\
70  &1.10   &0.662   &1.50   &0.0340     &3.7632     &0.123         &0.0592       &3.73      &117.87    &8.1555  &14.578  &144.61    \\
71  &1.10   &0.662   &1.50   &0.0340     &3.7630     &0.126         &0.0606       &3.80      &117.30    &8.0715  &14.408  &144.61    \\
72  &1.10   &0.662   &1.50   &0.0340     &3.7629     &0.128         &0.0620       &3.87      &116.74    &7.9327  &14.212  &144.61    \\
73  &1.10   &0.662   &1.50   &0.0340     &3.7628     &0.130         &0.0634       &3.94      &116.18    &7.8519  &14.055  &144.61    \\
74  &1.10   &0.662   &1.50   &0.0340     &3.7626     &0.132         &0.0648       &4.01      &115.62    &7.7468  &13.894  &144.62    \\
75  &1.10   &0.669   &1.00   &0.0340     &3.7602     &0.114         &0.0604       &4.15      &117.30    &\nodata &\nodata &144.51
\enddata
\tablecomments{The above mean large spacings were calculated averaging over \( l = 0, 1, 2, 3 \) and \(n = 10, 11, 12,...,30\).  The mean small spacings were averages over \(n = 10, 11, 12,...,30\) at a fixed {\it l} (as indicated)}
\end{deluxetable}

\clearpage

\end{document}